\documentclass[twocolumn,preprintnumbers, amssymb,amsmath,aps,floatfix,prl,nofootinbib,superscriptaddress,showpacs]{revtex4-1}

\usepackage{epsfig}
\usepackage{bm}
\usepackage{amssymb}
\usepackage{amsmath}
\usepackage{color}

\usepackage[colorlinks,
            linkcolor=blue,
            anchorcolor=black,
            citecolor=blue
            ]{hyperref}
\newcommand{\nn}{\nonumber \\}
\newcommand{\beq}{\begin{eqnarray}}
\newcommand{\eeq}{\end{eqnarray}}

\begin{document}
\title{Proton Spin Structure at Small-$x$}

\author{Renaud Boussarie}
\affiliation{Physics Department, Building 510A, Brookhaven National Laboratory, Upton, NY 11973}

\author{Yoshitaka Hatta}
\affiliation{Physics Department, Building 510A, Brookhaven National Laboratory, Upton, NY 11973}

\author{Feng Yuan}
\affiliation{Nuclear Science Division, Lawrence Berkeley National
Laboratory, Berkeley, CA 94720, USA}

\begin{abstract}
We generalize the Bartels-Ermolaev-Ryskin approach for the $g_1$ structure function at small-$x$~\cite{Bartels:1995iu,Bartels:1996wc} to determine the small-$x$ asymptotic behavior of the orbital angular momentum distributions in QCD.  We present an exact analytical solution of the evolution equation in the double logarithmic approximation and discuss its implications for the proton spin problem.  
\end{abstract}
\maketitle


{\it 1. Introduction.} Nucleon spin of $1/2$ is one of the fundamental properties of the building blocks of our universe. The spin sum rule describes how the proton's constituents contribute to its spin,
\beq
 \frac{1}{2}=\frac{1}{2}\Delta\Sigma + \Delta G +L_q +L_g, 
\eeq
including the quark/gluon helicity and orbital angular momentum (OAM) contributions \cite{Jaffe:1989jz}. The measurement of individual terms has been a major focus in experiments  at worldwide facilities, such as RHIC at BNL, the JLab 12 GeV upgrade, and the future Electron-Ion Collider (EIC). These experiments  probe the quark/gluon contributions at particular momentum fractions $x$ of the nucleon carried by the partons. There has been tremendous progress in constraining the quark/gluon helicity contributions from decades of experiments~\cite{deFlorian:2014yva,Nocera:2014gqa}, while, at the same time, the OAM contributions started to attract strong interest from both the theory and experimental communities \cite{Jaffe:1989jz,Lorce:2011kd,Hatta:2011ku,Lorce:2011ni,Ji:2012sj,Hatta:2012cs,Ji:2012ba}. 
 We anticipate great outcome from the planned EIC to finally answer this question on the nucleon spin sum rule~\cite{Boer:2011fh,Accardi:2012qut}. 
It is important to have a theoretical guidance for the small-$x$ behavior for the individual terms in the above sum rule. This is because any collider machine is limited by the kinematic reach in the small-$x$ part. Comparing the theoretical understanding of small-$x$ evolution with the experimental data will play a very important role in determining how small-$x$ we have to go before we can conclude the test of the spin sum rule. 

On the theory side, the small-$x$ evolution of spin distributions is one of the most intriguing questions in QCD. In the ordinary DGLAP approach to the parton helicity distributions,  the leading small-$x$ limit is obtained by the standard double logarithmic approximation (DLA) which resums powers of $(\alpha_s\ln(Q^2/\mu^2)\ln(1/x))^n$ at each order of perturbation theory \cite{Ahmed:1975tj,Ball:1995ye}. 
However, at very small-$x$, one finds a different kind of double logarithms $(\alpha_s \ln^2 1/x)^n$ whose resummation is highly nontrivial. This has been accomplished in  classic papers by Bartels-Ermolaev-Ryskin  (BER)  \cite{Bartels:1995iu,Bartels:1996wc} with the help of the so-called infrared evolution equation (IREE) \cite{Kirschner:1982qf,Kirschner:1983di}. 
More recently, Kovchegov-Pitonyak-Sievert (KPS) have explored this question from a different approach, by treating polarization effects as sub-eikonal corrections to the usual Wilson line formalism of high energy QCD \cite{Kovchegov:2015pbl,Kovchegov:2016weo,Kovchegov:2016zex,Kovchegov:2017jxc,Kovchegov:2017lsr,Kovchegov:2018znm,Kovchegov:2018zeq} (see, also, \cite{Altinoluk:2014oxa,Altinoluk:2015gia,Balitsky:2015qba,Balitsky:2016dgz,Hatta:2016aoc,Chirilli:2018kkw}). These developments have stimulated a lot of interest in the community. 

In this paper, we extend the original BER formalism to investigate the small-$x$ behavior for the quark/gluon OAM distributions and compare our result with the recent approaches based on DGLAP  \cite{Hatta:2018itc} and  the KPS formalism \cite{Kovchegov:2019rrz}. We first examine the anomalous dimensions associated with the OAM operators in the Wandzura-Wilczek (WW) approximation. This leads to remarkably simple relations between the OAM and helicity distributions at small-$x$. We then derive an extension of the BER evolution equations to include the OAM distributions. The solution of the extended BER equation predicts a behavior consistent with the one obtained from the anomalous dimension analysis. 
It has been known that, by perturbatively expanding the nonlinear BER evolution equations, one can obtain the leading small-$x$ DGLAP kernel at arbitrary higher orders~\cite{Bartels:1996wc,Blumlein:1996hb}. This has been explicitly confirmed up to three-loops~\cite{Moch:2014sna}, providing a crucial cross check for the BER derivation. Similarly, we expect the splitting kernel associated with the OAM distributions from our results can be checked by a future collinear computation. 

The rest of the paper is organized as follows. In Section 2, we apply the WW approximation and study the anomalous dimensions for the OAM distributions. In Section 3, we derive the BER evolution equation for all the four terms in the spin sum rule. The solution will be discussed in Section 4. By exploring the symmetry property of the splitting kernel we find a simple analytic solution for the OAM distributions at small-$x$ which depends on the quark/gluon helicity distributions solution of the original BER evolution equations. 
We then discuss the property of the solution and comment on its phenomenological consequences. Finally, we summarize our paper in Section 5.

{\it 2. Orbital angular momentum distributions in QCD.} Compared to their helicity counterparts, the Bjorken-$x$ distributions for the OAM of quarks $L_q(x)$ and gluons $L_g(x)$ are not commonly known. They have been first introduced in Refs.~\cite{Hagler:1998kg,Harindranath:1998ve} where their one-loop (`DGLAP') equation has also been discussed. However, the original definition of $L_{q,g}(x)$ is not gauge invariant, and the authors used the light-cone gauge for its physical interpretation as well as the actual calculations. More recently, the exact gauge invariant definition of $L_{q,g}(x)$ has been given in \cite{Hatta:2011ku,Ji:2012sj,Hatta:2012cs,Ji:2012ba}. Here we consider only the flavor singlet combination including the antiquark contribution: $L_q(x) = \sum_f (L_f(x) + L_{\bar{f}}(x))$.
A detailed analysis based on the QCD equations of motion and the Lorentz invariant relations shows that, unlike $\Delta\Sigma(x)$ and $\Delta G(x)$, $L_{q}(x)$ and $L_g(x)$ are not the usual twist-two parton distribution functions. Rather, they can be written as the sum of the Wandzura-Wilczek contribution and the genuine twist-three contribution~\cite{Hatta:2012cs}. The former is given by
   \beq
L_q^{(WW)}(x) &=& x\int^1_x \frac{dx'}{x'} (H_q(x')+E_q(x')) \nonumber\\
&&-x\int^1_x \frac{dx'}{x'^2}\Delta \Sigma(x')\ ,
\label{ev1}\\
L_g^{(WW)}(x)&=& x\int^1_x \frac{dx'}{x'} (H_g(x')+E_g(x')) \nn
&&-2x\int^1_x \frac{dx'}{x'^2}\Delta G(x') 
\label{ev2}
\eeq
where $H_{q,g}$ and $E_{q,g}$ are the standard twist-two generalized parton distributions in the limit of zero momentum transfer $\Delta\to 0$. 
Given these  relations, it is straightforward to write down the evolution equation for $L_{q,g}(x)$. To do so, we first define moments $L_{q,g}^\omega\equiv\int_0^1 dx x^{\omega-1}L_{q,g}(x)$  and write, e.g., (\ref{ev1}), 
in the moment (Mellin) space
\beq
L_q^\omega&=& \frac{1}{\omega+1}(H_q^{\omega+1} +E_q^{\omega+1}) -\frac{1}{\omega+1}\Delta\Sigma^\omega+\cdots, 
\label{coin}
\eeq 
where dots denote the contribution from the genuine twist-three distributions.  $H_{q}$ and $E_{q}$ evolve with the standard twist-two anomalous dimension matrix $\gamma_{ij}^\omega$ ($i,j=q,g$), and $\Delta \Sigma$ and $\Delta G$ evolves with the polarized anomalous dimensions $\Delta \gamma^\omega_{ij}$. We then immediately obtain
\beq
&&\frac{\partial}{\partial \ln Q^2} \begin{pmatrix} L^\omega_q \\ L_g^\omega \end{pmatrix} = 
\begin{pmatrix} \gamma^{\omega+1}_{qq} & \gamma^{\omega+1}_{qg} \\ \gamma^{\omega+1}_{gq} & \gamma^{\omega+1}_{gg} \end{pmatrix}  \begin{pmatrix} L^\omega_\Sigma \\ L_g^\omega \end{pmatrix}      +\frac{1}{\omega+1} \nn
&&~~~\times\begin{pmatrix} \gamma_{qq}^{\omega+1}- \Delta \gamma^{\omega}_{qq} & 2\gamma_{qg}^{\omega+1}- \Delta \gamma^\omega_{qg} \\ \gamma_{gq}^{\omega+1}-2\Delta \gamma^\omega_{gq} & 2\gamma_{gg}^{\omega+1}- 2\Delta \gamma_{gg}^\omega \end{pmatrix}  \begin{pmatrix} \Delta \Sigma^\omega \\ \Delta G^\omega \end{pmatrix}  +\cdots.
 \label{wa}
 \eeq
This is in agreement with the result of explicit one-loop calculations in \cite{Hagler:1998kg} which was later re-derived in \cite{Hoodbhoy:1998yb} in a heuristic way. As suggested in the latter paper, and as our derivation clearly indicates, this equation is valid only in the Wandzura-Wilczek approximation neglecting the genuine twist-three contributions. On the other hand, within this approximation it is valid to all orders in $\alpha_s$.  

We see that the small-$x$ behavior of $L_{q,g}(x)$ is determined by the $\omega\to 0$ behavior of the anomalous dimension matrix $\gamma^{\omega+1}_{ij}$ and $\Delta\gamma^\omega_{ij}$. In perturbation theory, the former involves a  single logarithmic series $(\alpha_s/\omega)^n$ and the latter involves a  double logarithmic series $(\alpha_s/\omega^2)^n$ which is more singular. We thus expect that the small-$x$ behavior of $L_{q,g}(x)$ are governed by  the helicity distributions. This has an immediate consequence for the relative prefactor between the helicity and OAM distributions.  Namely,  if $\Delta \Sigma(x)$ and $\Delta G(x)$ have a power-law behavior at small-$x$,
\beq
\Delta \Sigma(x) , \Delta G(x) \sim {1}/{x^\alpha}, \label{bee}
\eeq 
from (\ref{ev1}) and (\ref{ev2}) we obtain
\beq
L_q(x) \approx -\frac{1}{1+\alpha} \Delta \Sigma (x), ~ L_g(x) \approx -\frac{2}{1+\alpha}\Delta G(x). \label{em}
\eeq
The crucial minus sign indicates that there is a significant cancellation between the helicity and orbital angular momentum as first pointed out in \cite{Hatta:2016aoc} and repeatedly observed in \cite{Hatta:2018itc,More:2017zqp}. However, it disagrees with the recent result  $|L_g(x)|\ll |\Delta G(x)|$   in \cite{Kovchegov:2019rrz} based on the KPS approach.
We emphasize that (\ref{em}) is a robust prediction, in the sense that it has been derived from the exact QCD relations (\ref{ev1}) and (\ref{ev2}). In particular, it does not depend on the approximation one chooses to evaluate the anomalous dimension $\Delta \gamma^\omega_{ij}$. (This only affects the value of $\alpha$.) It can only be violated when one (or both) of  the two assumptions---the suppression of the genuine twist-three contribution and that of the single logarithmic contributions---turns out to be incorrect.


{\it 3. Infrared Evolution Equation for OAM distribution.} The relations (\ref{em}) dictate that the small-$x$ behavior of $L_{q,g}(x)$ should be the same as that of the corresponding helicity distributions. In order to determine the latter, one has to resum the  double logarithmic series $(\alpha_s \ln^2 1/x)^n$ (or $(\alpha_s/\omega^2)^n$ in the moment space) which appears in the perturbative calculation of the polarized splitting function $\Delta P(x)$ (or the anomalous dimension $\Delta \gamma^\omega$ in the Mellin space). This resummation has a long history. Problems show up already in QED in certain kinematical regimes of $e^+e^-$ scattering \cite{Gorshkov:1966ht} where one has to resum electron ladder diagrams with photon rungs. The situation in QCD is particularly challenging because the quark ladder and gluon ladder can mix under evolution. Moreover, these ladders are dressed up by the so-called Bremmstrahlung gluons which destroy the ladder structure. Thus the resummation of double logarithms is considerably harder than the usual BFKL resummation in unpolarized scattering where one only needs to consider gluon ladder diagrams. Nevertheless, the formalism to tackle this problem, the  Infrared Evolution Equation (IREE), 
is well established in the literature mostly due to Kirschner and Lipatov \cite{Kirschner:1982qf,Kirschner:1983di,Ermolaev:1995fx,Bartels:1995iu,Kirschner:1996jj,Ermolaev:2007yb}, culminating in the determination by BER~\cite{Bartels:1996wc}  of the exponent $\alpha$ in (\ref{bee}) for the helicity distributions. 
In  the following, we shall demonstrate that IREE can be generalized to include the OAM distributions, and explicitly solve this equation.

The double logarithm $\alpha_s \ln^2 (1/x)$ at each order in  perturbation theory comes from the region of phase space where the {\it lifetime} of partons is strongly ordered 
\beq
&&\frac{\beta_1}{k^2_{1\perp} }\gg \frac{\beta_2}{k^2_{2\perp}} \gg \cdots \gg \frac{\beta_n}{k^2_{n\perp} }, \nonumber \\ 
&&\beta_1\gg \beta_2\gg \cdots \gg \beta_n, \qquad  k_{i\perp}^2\gg\mu^2
\eeq
where $\beta_i$ and $k_{i\perp}$ are the longitudinal and transverse components of $i$th parton in the ladder configuration.  As the name suggests, IREE is an evolution equation in the infrared cutoff scale $\mu^2$ (or its conjugate variable $\omega$ in the Mellin space). The crucial observation is that since the transverse momenta are not strongly ordered, $k_{i\perp}^2$ with any $i$ can be the softest momentum along the ladder bounded from below by $\mu^2$. By keeping track of this $\mu^2$-dependence, one can recover the $x$ and $Q^2$ dependence of the structure function since the latter can be viewed as a function of $s/\mu^2= Q^2/(\mu^2 x)$ and $Q^2/\mu^2$.

Let us introduce a short-hand notation to represent the four-component spin density vector: ${\rm\bf S}(x,Q^2)=\left(\Delta \Sigma(x,Q^2),\Delta G(x,Q^2), L_q(x,Q^2),L_G(x,Q^2)\right)$. The evolution equation is normally expressed in the moment space, 
\beq
{\rm\bf S}(x,Q^2)=\int \frac{d\omega}{2\pi i } \left(\frac{1}{x}\right)^\omega {\rm\bf S}^\omega(Q^2) \ ,
\eeq
and the evolution equation takes the form,
\beq
\frac{\partial}{\partial \ln Q^2}{\rm\bf S}^\omega(Q^2)=\frac{1}{8\pi^2}F_0{\rm\bf S}^\omega(Q^2) \ .
\eeq
In the above equation, the $4\times 4$ matrix $F_0$ represents the splitting kernel for the spin density distributions at small-$x$,
\beq
F_0(\omega)=\begin{pmatrix}  F_{qq} & F_{qg} & 0 & 0 \\ F_{gq} & F_{gg} & 0 & 0 \\ F_{L_q q}& F_{L_q g} & F_{L_q L_q} & F_{L_q L_g} \\
F_{L_g q} & F_{L_g g} & F_{L_g L_q} & F_{L_g L_g} \end{pmatrix}\ ,
\eeq
where the subscript 0 denotes `color-singlet'. This is a generalization of the $2\times 2$ matrix in the BER derivation where $F_{ij}(\omega)$ is identified with the near-forward scattering amplitude between partons $i$ and $j$ in the Mellin space. 
The new components related to OAMs may be difficult to interpret as  `scattering amplitude' but we can still interpret them as anomalous dimensions.   The upper-right corner of $F_0$ is zero because the evolution of helicity distributions, being purely twist-two objects, is not affected by the OAM distributions. (Differently from BER, we rearrange the matrix elements in the `normal' order in which the quark  appears  in the first row. We think most readers are used to this notation.) $F_0$ satisfies the following recursion relation 
  \beq
F_0 = \frac{g^2}{\omega}M_0 -\frac{g^2}{2\pi^2\omega^2}F_8 G_0 + \frac{1}{8\pi^2\omega}F_0^2, \label{ber1}
\eeq
where ($T_f\equiv n_f/2$)
\beq
M_0 = \begin{pmatrix} C_F & -2T_f & 0 & 0 \\ 2C_F & 4C_A & 0 & 0 \\ -C_F & 2T_f & 0 & 0 \\ -2C_F & -4C_A & 2C_F & 2C_A \end{pmatrix} \ .
\eeq
The above equation is coupled with $F_8$, the scattering amplitude matrix with octet color exchange in the $t$-channel. The constant matrix 
\beq
G_0 = \begin{pmatrix} C_F & 0 & 0 & 0 \\ 0 &  C_A & 0  & 0 \\ 0 & 0 &  C_F & 0 \\ 0 & 0 & 0 & C_A 
\end{pmatrix}.
\eeq
represents the modification in color factors due to the presence of a soft Bremsstrahlung gluon. $F_8$ satisfies a closed equation \cite{Kirschner:1983di}
\beq
F_8 = \frac{g^2}{\omega}M_8 + \frac{g^2C_A}{8\pi^2\omega} \frac{d}{d\omega}F_8 + \frac{1}{8\pi^2 \omega}F_8^2. \label{b3}
\eeq
where 
\beq
M_8 = \begin{pmatrix} -1/2N_c & -T_f  & 0 & 0 \\ C_A & 2C_A & 0 & 0 \\ 1/2N_c & T_f & 0 & 0 \\ -C_A & -2C_A & C_A & C_A  \end{pmatrix} ,
\eeq
is the splitting kernel in the octet channel. In Appendix A, we explain how to obtain $M_8$ and $G_0$. 

The upper-left $2\times 2$ matrices in $M_0$, $M_8$, and $G_0$ are the same as those in the BER paper. The rest are the new results in our paper. For $M_0$,  the lower-left $2\times 2$ matrix is the same as the upper-left $2\times 2$ matrix  but with an opposite sign. This is related to the angular momentum conservation at the one-loop level. The same should be true in $M_8$ because the difference in color factors does not modify this kinematical effect, and this explains the lower-left $2\times 2$ matrix in $M_8$.  What is not so obvious is the lower-right $2\times 2$ matrix of $M_8$. This has been obtained by comparing the second row of $M_0$ and $M_8$. When switching from color singlet to color octet, the second row changes as $(2C_F,4C_A) \to (C_A,2C_A)$. That is, in the $gq$ channel $2C_F$ turns into $C_A$ and in the $gg$ channel the coefficient is halved. We have implemented the same change in the $gq$ and $gg$ channels of the OAM, that is, $(2C_F,2C_A) \to (C_A,C_A)$. This simple prescription is justified because the color factor should be the same when calculating, for example, the $L_q\to L_g$ and $q\to g$ splittings.

As already observed by BER and elaborated in \cite{Blumlein:1996hb}, solving IREE is equivalent to perturbatively resumming  the anomalous dimension matrix $\Delta \gamma^\omega$ to all orders in $(\alpha_s/\omega^2)^n$. As a matter of fact, the matrix $F_0(\omega)$ is directly proportional to $\Delta \gamma^\omega$. To explain this, here we show the iterative solution  of (\ref{ber1}) and (\ref{b3}) to ${\cal O}(g^6)$ 
\beq
F_0(\omega)&=& \frac{g^2}{\omega} M_0 +\frac{g^4}{8\pi^2\omega^3}(M_0^2-4M_8G_0) \nn
&& + \frac{g^6}{32\pi^4\omega^5} \left(M_0^3+2C_A M_8G_0 -2M_8^2G_0\right.\nn
&&\left.-2M_0M_8G_0 -2M_8G_0M_0 \right) . \label{ite}
 \eeq	
Via the inverse Mellin transform, we obtain the most singular part of the three-loop splitting function ($a_s =\frac{\alpha_s}{4\pi}$) \cite{Blumlein:1996hb}
 \beq
\Delta P(x)&=& {F_0(x)}/{8\pi^2}\nn
&=& a_s 2M_0 + a_s^2 \ln^2\frac{1}{x} 2 (M_0-4M_8G_0) \nn
 && + a_s^3 \ln^4\frac{1}{x} \frac{2}{3}\left(M_0^3+2C_A M_8G_0 -2M_8^2G_0 \right.\nn
 &&\left.-2M_0M_8G_0 -2M_8G_0M_0\right) .\label{three}
 \eeq
This agrees with the result of explicit three-loop calculations of the splitting function in 2014  \cite{Moch:2014sna}. That is, BER's paper in 1996 has correctly predicted the small-$x$ limit of this three-loop result.


{\it 4. Solution of IREE for the OAM.} In the $2\times 2$ case, Eq.~(\ref{b3}) can be solved analytically, and after the solution is substituted into (\ref{ber1}), the resulting equation can be solved numerically, or analytically under certain approximations. What happens is that usually  the resulting function $F_0(\omega)$ has singluarities (branch cut) in the complex $\omega$-plane. The rightmost singularity at $\omega=\omega_s$ then determines the small-$x$ exponent $\alpha=\omega_s$ in (\ref{bee}).

While it is straightforward to numerically solve the $4\times 4$ version of IREE,   remarkably one can derive   an exact analytical solution via the following heuristic argument.  Consider the ${\cal O}(g^4)$ (two-loop) term of the iterative solution (\ref{ite}) 
 \beq
&&M_0^2-4M_8G_0 = \\
&&\begin{pmatrix} C_F^2 + \frac{2C_F}{N_c}-4C_FT_f & -4C_AT_f-2C_F T_f & 0 & 0 \\ 
4C_A C_F +2C_F^2 & 8C_A^2-4C_F T_f & 0 & 0 \\ -C_F^2 - \frac{2C_F}{N_c}+4C_FT_f & 4C_AT_f+2C_F T_f & 0 & 0 \\
-8C_A C_F -4C_F^2 & -16C_A^2+8C_F T_f & 0 & 0 \end{pmatrix}\ .\nonumber \label{lower}
\eeq
The upper-left corner is the two-loop anomalous dimension known in the literature. We immediately notice that  the third and fourth columns are zero.  Moreover, the third row is $-1$ times  the first row, and the fourth row is $-2$ times the second row.   We have checked that exactly the same pattern appears in the ${\cal O}(g^6)$ (three-loop) solution. Assuming this to be true to all orders, we can immediately write down the exact solution of the $4\times 4$ IREE. Let  
\beq
F_0^{2\times 2}=\frac{g^2}{\omega} M_0^{2\times 2} + \begin{pmatrix} A_1 & A_2 \\ B_1 & B_2 \end{pmatrix},
\eeq
be the solution in the helicity part alone, that is, the BER solution. Then the full solution is simply 
\beq
F_0^{4\times 4}=\frac{g^2}{\omega} M_0^{4\times 4} + \begin{pmatrix} A_1 & A_2 & 0 & 0  \\ B_1 & B_2 & 0 & 0 \\ -A_1 & -A_2 & 0 & 0 \\ -2B_1 & -2B_2 & 0 & 0 \end{pmatrix}. \label{comp}
\eeq
A short proof of the above solution is provided in the Appendix B. We add that we have also confirmed this by numerically solving the $4\times 4 $ IREE. The solution of the renormalization group equation for the spin density distribution is then given by
\beq
{\rm\bf S}(x,Q^2)=\int \frac{d\omega}{2\pi i } \left(\frac{1}{x}\right)^\omega \left(\frac{Q^2}{\mu^2}\right)^{\frac{F^{4\times 4}_0}{8\pi^2}}{\rm\bf S}^\omega(\mu^2) \ .
\eeq

It is important to notice that 
the factors $-1$ and $-2$ in the third and fourth rows of (\ref{comp}) can be recognized in the coefficients of $\Delta \gamma^\omega$ in (\ref{wa}). Actually, the structure (\ref{comp}) is precisely what one expects from the full equation (\ref{wa}) in the double log approximation. 
Indeed,  $\gamma^{\omega+1}, \Delta \gamma^\omega \sim \alpha_s/\omega$ at one-loop in DLA, and these are collected in the first term of (\ref{comp}). At $n$-loop $(n>1)$, one only keeps $\Delta\gamma^\omega \sim \alpha_s/\omega (\alpha_s/\omega^2)^{n-1}$ and set $\gamma^{\omega+1}$ to be zero.  This is the second term of (\ref{comp}). (We also need to approximate $1/(\omega+1) \approx 1$ since $\omega$ is formally small.) Therefore, our solution  is fully consistent with the result from the QCD equation of motion  (\ref{wa}). While this should be the case, given the complexity of the equation it is highly nontrivial that the straightforward generalization of IREE to the OAM sector automatically satisfies this constraint.  

 IREE allows one to determine the small-$x$ exponent $\alpha$ as well as the proportionality constant between $L_q(x)$ and $L_g(x)$ about which the equation of motion relation (\ref{em}) has nothing to say. 
We can further diagonalize 
Eq.~(\ref{comp}) at the BER saddle point $\omega= \omega_s= 3.45\sqrt{\frac{\alpha_sN_c}{2\pi}}$ (for $n_f=4$ flavors). This gives four eigenvectors  
\beq
S_1(x)&\approx & 0.29\Delta \Sigma(x) + \Delta G(x), \nn
S_2(x)&\approx &2.29 \Delta \Sigma(x)  +\Delta G(x),  \nn
  S_3(x)&=&\frac{C_F}{N_c}(\Delta \Sigma (x)+L_q(x)) +  2\Delta G(x)+L_g(x), \nn
 S_4(x)&=&\Delta \Sigma(x) +L_q(x),
\eeq
 where we set $\alpha_s=0.18$ so that $\omega_s=1.01$.  $S_1(x)$ corresponds to the largest eigenvalue of $F_0$. 
Requiring that $S_{2,3,4}(x)$ are subleading at small-$x$ and large-$Q^2$, we  arrive at the relations
\beq
&&\Delta G(x) \approx -2.29\Delta \Sigma(x) \propto  x^{-3.45\sqrt{\frac{\alpha_sN_c}{2\pi}}}\sim 
\frac{1}{x^{1.01}}, \nn
&&
L_g(x)\approx -2\Delta G(x),~~ \Delta \Sigma(x) \approx - L_q(x).
 \label{final}
\eeq
These are the main results from our derivations. The relative coefficients between the helicity and OAM distributions agree with  the independently obtained result (\ref{em}) in the formal DLA limit $\alpha\ll 1$ (see also \cite{Hatta:2016aoc,Hatta:2018itc}), though in practice $\alpha=\omega_s$ is numerically close to, or even exceeds unity in the BER solution.  This seems worrisome because $\alpha>1$ leads to diverging  first moments $\Delta \Sigma=\int_0^1dx\Delta \Sigma(x)$ etc. However,  various corrections such as the running coupling effect \cite{Ermolaev:2003zx}, subleading logarithms \cite{Ermolaev:2009cq} and nonperturbative effects (see, e.g., \cite{Hatta:2009ra}) will bring down the value of $\alpha$.\footnote{We note that $\alpha=\omega_s$ can be made slightly smaller by increasing $n_f$. For example,  with $n_f=5$ which may be more appropriate for $\alpha_s=0.18$, we find $\omega_s=0.99$. }

{\it 5. Summary.} In this paper, we have investigated the small-$x$ behavior for the quark/gluon OAMs based on leading double logarithmic ($\alpha_s\ln^2(1/x)$) resummation formalism of BER approach. From the solutions of the relevant evolution equation, we have found that the OAM distributions have the same power behavior as  their helicity counterparts.  

As already commented in \cite{Hatta:2016aoc,Hatta:2018itc}, the relative negative sign between the helicity and OAM distributions, in both the quark and gluon sectors, is phenomenologically important since the current best estimate of $\Delta G$ suffers from large uncertainties  from the small-$x$ region \cite{deFlorian:2014yva}, and reducing these uncertainties is one of the goals of  the future EIC. While of course the precise value of $\Delta G$ is a fundamental question of QCD,  our result suggests that the resolution of the nucleon spin puzzle does not reside in the helicity distributions at small-$x$ where they are canceled by the OAM distributions, but should be looked for in the large-$x$ region of the OAMs.   Proposal have been made to experimentally access $L_{q,g}(x)$ in the medium to large-$x$ region \cite{Ji:2016jgn,Bhattacharya:2017bvs,Bhattacharya:2018lgm} and small-$x$ region \cite{Hatta:2016aoc}, but we think more theoretical effort in this direction is highly needed.

Finally, it is a challenging problem to include the GPD contributions $H_{q,g}$ and $E_{q,g}$ in (\ref{ev1}) and (\ref{ev2}). For this purpose, one has to go beyond DLA and resum also single (BFKL) logarithms.  At very small-$x$, one may also have to include  the gluon saturation (higher twist) effects.
 A promising approach toward these goals is the Wilson line formalism of high energy QCD with sub-eikonal corrections \cite{Kovchegov:2015pbl,Kovchegov:2016weo,Kovchegov:2016zex,Kovchegov:2017jxc,Kovchegov:2017lsr,Kovchegov:2018znm,Kovchegov:2018zeq,Altinoluk:2014oxa,Altinoluk:2015gia,Balitsky:2015qba,Balitsky:2016dgz,Hatta:2016aoc,Chirilli:2018kkw,Tarasov:2019rfp}, although it remains to be seen how one can first recover the BER result in this framework.   Another interesting problem is to include the contribution from the genuine twist-three distributions (neglected in (\ref{coin}), (\ref{wa})). Progress in this direction is underway, and will be reported  elsewhere.

\acknowledgements
We thank Yuri Kovchegov for many stimulating discussions. We also thank  Werner Vogelsang and Bowen Xiao for discussions.
This material is based upon work supported by the LDRD programs of 
Lawrence Berkeley National Laboratory and Brookhaven National Laboratory, the U.S. Department of Energy, 
Office of Science, Office of Nuclear Physics, under contract numbers 
DE-AC02-05CH11231 and  DE-SC0012704.    It is also supported by the Natural Science Foundation of China (NSFC) under Grant No.~11575070.

\onecolumngrid
\newpage
 
{\bf Supplemental  material}
\appendix

\section{Appendix A: Color factors in the octet channel}

In this Appendix, we reproduce the color factors associated with the matrices $M_8$ and $G_0$ which BER presented without derivation.  A color singlet quark ladder exchange can be viewed as  quark-antiquark scattering in the $t$-channel. We use the following  projectors  of their color indices 
\beq
\Gamma_{ij} = \frac{1}{N_c}\delta_{ij}, \qquad \Gamma^a_{ij} = \frac{2}{N_c^2-1}t^a_{ij},
\eeq
 where $i,j=1,2,...,N_c$ and $a=1,2,...,N_c^2-1$, for the singlet and octet channels, respectively. The normalization is determined by contracting with $\delta_{ij}$ and $t^a_{ij}$, respectively. 
Similarly, for gluon-gluon scattering, we use
\beq
\Gamma_{ab}=\frac{1}{N_c^2-1}\delta_{ab}, \qquad \Gamma^a_{bc}=\frac{-if_{abc}}{N_c(N_c^2-1)},
\eeq
with the normalization determined by contracting with $\delta_{ab}$ and $T^a_{bc}=-if_{abc}$. With these projectors, the color factors can be easily calculated. 
For $q\to q$ splitting in the singlet channel (one-gluon exchange between a color singlet $q\bar{q}$ pair), we have 
\beq
\frac{1}{N_c}{\rm Tr}[t^at^a] =C_F.
\eeq
This is just the usual color factor in the one-loop DGLAP splitting function. In the octet channel, we get
\beq
\frac{2}{N_c^2-1}{\rm Tr}[t^a t^bt^a t^b] = -\frac{1}{2N_c}. \label{qq}
\eeq
 This explains the $qq$ component of $M_8$. 
For $g\to g$ splitting in the color singlet channel,
\beq
\frac{1}{N_c^2-1}f_{abc}f_{abc}= C_A,
\eeq
while in the color octet channel, 
\beq
\frac{1}{N_c(N_c^2-1)}f_{abc}f_{ade}f_{gbd}f_{gce}=\frac{C_A}{2}. \label{oc}
\eeq
Compared to the $gg$ component of $M_0$ and $M_8$, there is a factor 4 difference. This comes from the Lorentz indices as seen in Eq.~(2.25) of \cite{Bartels:1996wc}.  The present discussion can only determine the relative factor between $M_0$ and $M_8$.  
For $q\to g$ splitting in the color singlet channel,
\beq
\frac{1}{N_c}{\rm Tr}[t^a t^a]=C_F,
\eeq
and in the color octet channel,
\beq
\frac{2}{N_c^2-1}{\rm Tr}[t^a t^b t^c] (-if_{abc})=\frac{C_A}{2}. \label{qg}
\eeq 
Finally for $g\to q$ splitting in the singlet channel
\beq
\frac{1}{N_c^2-1}{\rm Tr}[t^a t^a] = \frac{1}{2},
\eeq
while in the octet channel,
\beq
\frac{-if_{abc}}{N_c(N_c^2-1)}{\rm Tr}[t^a t^bt^c]=\frac{1}{4}. \label{gq}
\eeq

We also explain the role of the matrix $G_0$. This is the color factor needed to convert  the splitting in the octet channel into the splitting in the singlet channel by attaching  a Bremsstrahlung gluon loop. 
The color factor of the latter is, in the $q\to q$ channel,
\beq
\frac{1}{N_c}{\rm Tr}[t^a t^b t^a t^b]= \frac{-1}{2N_c}C_F.
\eeq
Compared to (\ref{qq}), there is a relative factor $C_F$. Similarly, in the $q\to g$ channel, the singlet channel with one gluon loop gives 
\beq
\frac{1}{N_c} {\rm Tr}[t^a t^b t^c] (-if_{abc}) =\frac{N_c^2-1}{4}.
\eeq
This is $C_F$ times (\ref{qg}). Therefore, the first entry of $G_0$ is $C_F$, representing the emission of a Bremsstrahlung gluon from a quark.

On the other hand, in the singlet $g\to g$ channel with one Bremsstrahlung gluon, the color factor is 
\beq
\frac{1}{N_c^2-1}f_{abc}f_{cde}f_{bgd}f_{gae}=\frac{C_A^2}{2}.
\eeq
Compared to (\ref{oc}), we get a factor $C_A$. Similarly, in the $g\to q$ channel 
\beq
\frac{1}{N_c^2-1}{\rm Tr}[t^at^bt^c] (-if_{abc}) = \frac{C_A}{4},
\eeq
which is $C_A$ times  (\ref{gq}). Therefore, the second entry of $G_0$ is $C_A$, representing the emission of a Bremsstrahlung gluon from a gluon.

At this point one may  wonder why only $C_A$ appears in the second term of (\ref{b3}). This is because $F_8$ is a signature-even amplitude, and in this case one finds the factor $C_A$ even when the Bremsstrahlung gluon is emitted from a quark, see Eq.~(3.13) of  \cite{Kirschner:1983di}.

\section{Appendix B: A short proof of Eq.~(\ref{comp})}

Here we provide a short proof of the solution for $F_0$ in Eq.~(\ref{comp}). To check this is indeed the solution, notice that 
\beq
M^{4x4}_0 = E \begin{pmatrix} \hspace{2mm}\huge{M}_0^{2x2}&   \begin{matrix} 0 \hspace{3mm}& 0 \\ 0 \hspace{3mm}& 0 \end{matrix} \\ \begin{matrix} 0 & 0 \\   0& 0 \end{matrix} &
\begin{matrix} 0 \!\!& \!\!0 \\ 2C_F\! &\! 2C_A \end{matrix} \end{pmatrix}E^{-1}, \qquad M^{4x4}_8 = E \begin{pmatrix} \hspace{2mm}\huge{M}_8^{2x2}&   \begin{matrix} 0 \hspace{3mm}& 0 \\ 0 \hspace{3mm}& 0 \end{matrix} \\ \begin{matrix} 0 & 0 \\   0& 0 \end{matrix} &
\begin{matrix} 0 & 0 \\ C_A&C_A \end{matrix} \end{pmatrix}E^{-1},
\eeq
\beq
\begin{pmatrix} A_1 & A_2 & 0 & 0  \\ B_1 & B_2 & 0 & 0 \\ -A_1 & -A_2 & 0 & 0 \\ -2B_1 & -2B_2 & 0 & 0 \end{pmatrix} = E \begin{pmatrix} A_1 & A_2 & 0 & 0  \\ B_1 & B_2 & 0 & 0 \\ 0 & 0 & 0 & 0 \\ 0 & 0 & 0 & 0 \end{pmatrix}E^{-1}, \qquad G_0 =E G_0E ^{-1},
\eeq
where 
\beq
E= \begin{pmatrix} 1 & 0 & 0 & 0 \\ 0 & 1 & 0 & 0 \\ -1 & 0 & 1 & 0 \\ 0 &-2 & 0 & 1 \end{pmatrix} , \qquad E^{-1}=\begin{pmatrix} 1 & 0 & 0 & 0 \\ 0 & 1 & 0 & 0 \\ 1 & 0 & 1 & 0 \\ 0 &2 & 0 & 1 \end{pmatrix}.
\eeq
That is, all the matrices can be block-diagonalized by the same, constant matrix $E$. Once this is done,  the upper-left $2\times 2$ matrix problem is exactly the $2\times 2$ IREE solved by BER. 
On the other hand, the lower-right (LR) $2\times 2$ matrix problem reduces to the equations 
  \beq
F^{LR}_0 = \frac{g^2}{\omega}\begin{pmatrix} 0 & 0 \\ 2C_F & 2C_A \end{pmatrix}  -\frac{g^2}{2\pi^2\omega^2}F^{LR}_8 G_0^{2x2} + \frac{1}{8\pi^2\omega}(F^{LR}_0)^2, \label{bbbb}
\eeq
and in the octet sector, 
\beq
F^{LR}_8 = \frac{g^2}{\omega}  \begin{pmatrix} 0 & 0 \\ C_A & C_A \end{pmatrix} + \frac{g^2C_A}{8\pi^2\omega} \frac{d}{d\omega}F^{LR}_8 + \frac{1}{8\pi^2 \omega}(F^{LR}_8)^2.  \label{exact}
\eeq
(\ref{exact}) can be solved exactly as 
\beq
F^{LR}_8 = \frac{C_Ag^2}{\omega}\begin{pmatrix} 0 & 0 \\ 1 & 1 \end{pmatrix} . \label{f8}
\eeq
 (\ref{comp}) means that the solution of (\ref{bbbb}) must be
\beq
F^{LR}_0 = \frac{g^2}{\omega} \begin{pmatrix} 0 & 0 \\ 2C_F & 2C_A \end{pmatrix}. \label{f0}
\eeq
Indeed, when (\ref{f8}) and (\ref{f0}) are substituted into (\ref{bbbb}), the last two terms on the right hand side cancel exactly. This completes the proof that (\ref{comp}) is the solution.

\end{document}